\documentclass{article}
\usepackage{spconf,amsmath,graphicx,subfigure}
\usepackage{color, multirow, url}
\usepackage[breaklinks]{hyperref}
\usepackage{breakurl}

\title{Deep Autotuner: A Data-Driven Approach to Natural-sounding Pitch Correction for Singing Voice in Karaoke performances}

\name{Sanna Wager$^{1}$, George Tzanetakis$^{2, 3}$, Cheng-i Wang$^{3}$, Lijiang Guo$^{1}$, Aswin Sivaraman$^{1}$, Minje Kim$^{1}$\thanks{The research work done for this paper was supported by the internship program at Smule, Inc., in collaboration with the audio/video team.}}

\address{$^1$ Indiana University, School of Informatics, Computing, and Engineering, Bloomington, IN, USA\\
$^2$University of Victoria, Department of Computer Science, Victoria, BC, Canada\\
$^3$ Smule, Inc, San Francisco, CA, USA}

\begin{document}
\ninept
\maketitle
\begin{abstract}
%George:
%Repeat the need to get rid of the dependency on note boundaries as
%necessary for a developing a fully unsupervised auto-tuner later
%in the last section.
We describe a machine-learning approach to pitch correcting a solo singing performance in a karaoke setting, where the solo voice and accompaniment are on separate tracks. The proposed approach addresses the situation where no musical score of the vocals nor the accompaniment exists: It predicts the amount of correction from the relationship between the spectral contents of the vocal and accompaniment tracks. Hence, the pitch shift in cents suggested by the model can be used to make the voice sound in tune with the accompaniment. This approach differs from commercially used automatic pitch correction systems, where notes in the vocal tracks are shifted to be centered around notes in a user-defined score or mapped to the closest pitch among the twelve equal-tempered scale degrees. We train the model using a dataset of 4,702 amateur karaoke performances selected for good intonation. We present a Convolutional Gated Recurrent Unit (CGRU) model to accomplish this task. This method can be extended into unsupervised pitch correction of a vocal performance---popularly referred to as autotuning.
\end{abstract}
\begin{keywords}
music information retrieval, pitch, singing voice, automatic pitch correction, deep learning, autotuning
\end{keywords}
\section{Introduction}
\label{sec:intro}
Automatic singing pitch correction is a commonly desired feature in karaoke. Modifying a singer's pitch track to make it sound more in tune but not unnatural is not straightforward. In some cases, the sung melody is unknown, i.e. no musical score is linked to the performance. However, even without such {\em a priori} score information, a listener with a basic level of practice in music can often detect notes that are out of tune and predict which direction the singer should shift the pitches to make them sound more in tune only based on the level of perceived musical harmony. In this paper we envision an automatic pitch correction program that behaves similarly. To the best of our knowledge, the proposed method is the first data-driven approach to correcting singing voice pitch based on the accompaniment.

A fully automatic pitch correction---``autotuning"---algorithm is difficult to define. When a singer performs a melody, we assume that they have the sequence of notes in mind that define this melody---a musical score---whether or not this is notated on paper. A score is typically represented as a simple sequence or piano roll of note-wise constant pitches. Thus, the score only contains some of the information of a performance, and nuances are left to the singer. Indeed, a vocalist's pitch track follows the general contour of the score but varies continuously due to expressive gestures such as pitch bending, vibrato, and the physicality of the voice. Pitch variation techniques themselves differ based on musical genre, local musical context, and personal style. Within one performance, different levels of sharpness or flatness from note to note (i.e. deviations in pitch above or below the musical score), make it difficult to know exactly what pitch the singer intended. The task of shifting the singer's pitch so that it follows the general contour indicated by the musical score, adjusting unintended out-of-tune pitches without preserving the intentional variations, is not obvious. The proposed data-driven approach tries to respect the nuanced variations of sung pitch while the system also actively estimates the amount of unintended pitch shift.

\section{Related Work}

\subsection{Related pitch-correction systems}
The first commercial pitch-correction technique, Antares Auto-Tune \cite{antares:2016}, is also one of the most commonly used. It measures the fundamental frequency of the input monophonic singing recording, then re-synthesizes the pitch-corrected audio signal. In Auto-Tune and in recent work on continuous score-coded pitch correction \cite{salazar2015continuous}, the vocals can either be tuned automatically or manually. In the automatic case, each vocal note is pitch shifted to the nearest note in a user-input set of pitches (scale) or to the pitch in the score if it is known. In the manual case, a recording engineer uses the plugin's interface to move each note to the desired score and precise pitch. With either approach, the default musical scale is the equal-tempered scale, in which each pitch $p$ belongs to the set of MIDI pitches $[0, 1, ..., 127]$ and its frequency in Hertz is defined as $440*2^{\frac{p-69}{12}}$. Some users prefer a finer resolution and include more than twelve pitches per octave, or use intervals of varying sizes between pitches. In all cases, the continuous parameter, frequency, is discretized to a small set of values and the performance frequency of every note is shifted after correction to be centered exactly around its corresponding discrete $p$, which may not always be the most artistically desirable musical choice. This approach maps performance pitch to the musical score, thus not directly taking into account a singer's intentional pitch variation for expressive means. To avoid flattening the singer's pitch to the note, producing a robotic sound, a user-adjustable "time-lag" parameter that corrects pitch gradually introduces a tradeoff between preservation of pitch variation and accuracy. Furthermore, while the original Auto-Tune is most suitable for music following the Western twelve-tone scale or other known scales, it is not easily adaptable to other musical-cultural contexts with different scales or more fluidly varying pitch. Our proposed model focuses on the automatic approach in the case where a score is not present, representing pitch as a continuous instead of discrete parameter.

\subsection{Musical intonation studies on center pitch}
Pitch correction systems such as those described above assign to every note a center frequency around which all pitch variations are centered. Quantitative and qualitative studies on musical intonation show that professional-level singers and instrumentalists often center their frequencies at values that deviate from the equal-tempered scale. In particular, they often sing or play sharp relative to an accompaniment. This phenomenon is described in \cite{parncutt2018psychocultural}, but it dates back to \cite{barbour1938just}'s work on string instruments as well as work from the early twentieth century \cite{schoen1926pitch, cameron1907tonal}. Furthermore, soloists often center their singing at a higher frequency than the accompaniment, possibly in order to stand out \cite{kantorski1986string} \cite{rakowski1985perception}. Devaney {\it et al.} \cite{devaney2011intonation} measure much variety in musical interval sizes both above and below the equal-tempered intervals in the context of melodic intervals---where pitches are sequential in time---and polyphonic choral music performed by professionals and semi-professionals. In particular, ascending melodic intervals tend to be large, while descending intervals tend to be small. Another relevant result in \cite{parncutt2018psychocultural} is that frequency and perceived pitch are often slightly different. In the proposed system, we do assign one center frequency to every note as the target of the learning system, but the system lets the fundamental frequency take any value instead of belonging to a scale. The choice of center frequency is trained by applying a deep learning system to a dataset of performances, where singers took musical choices that sounded good to them, thus taking perceived pitch into account.

\subsection{Style transfer techniques}
Recent work transfers features from a professional-level performance to an amateur performance of the same song after first aligning the two performances in time. Luo et al. proposed to match the pitch contour of the professional-level performance while preserving the spectral envelope and aperiodicity of the amateur performance \cite{luo2018singing}. Meanwhile, Yong and Nam proposed to match both the pitch and amplitude envelopes \cite{yong2018singing}. Our model shares in common with these approaches the fact that it uses features gathered from high-level performances \cite{wager2018intonation, damp2018}. It differs from the previous work by not mapping style from one specific performance of the same song, instead, taking a data-driven approach to learn a supervised model that adjusts the center pitch of an unseen performance while preserves the original singer's style and characteristics.

\subsection{Deep learning research in music signal processing}
Gomez {\it et al.} \cite{gomez2018deep} describe recent work on deep learning for singing processing. More generally, in both music and speech, various combinations of recurrent convolutional neural networks have been successfully adopted in audio signal processing and Music Information Retrieval (MIR) applications. \cite{boulanger2012modeling} applies RNNs coupled with restricted Bolzmann machines to polyphonic pitch transcription. In \cite{basaranmain}, a Convolutional Gated Recurrent Unit (CGRU), where the GRU \cite{chung2014empirical} structure, an adaptation of RNNs that addresses the gradient vanishing problem, estimates the main melody in polyphonic audio signals pre-processed using the Constant-Q Transform (CQT) followed by Nonnegative Matrix Factorization \cite{LeeDD2000nips}. The convolutional layer structure is based on \cite{bittner2017deep}, a model for polyphonic pitch transcription that takes as input the six-dimensional harmonic CQT (HCQT) and outputs a two-dimensional transcription of the same shape as the input CQT, but with a single channel. Another example of CGRU can be found in \cite{xu2017convolutional} for sound event detection. The authors tested their model with three different input formats, Short-Time Fourier Transform (STFT), Mel-frequency bins, and raw audio, and reported best results using the Mel-Frequency Cepstrum Coefficient (MFCC) bins. The MFCC bins give the advantage over raw audio of being pre-processed in a structurally meaningful way, and log-spaced bins in MFCC present the advantage over the STFT bins of being translation invariant in the frequency domain, which makes them suitable for models involving convolutional layers. In detection problems, frequency resolution is not as important in detection as in problems involving pitch, where a few cents make a significant difference in performance. For this reason, we choose to use the CQT---which is also log-spaced but high-dimensional--- combined with a GRU. Our problem differs from pitch transcription problems, which predict pitch presence at a resolution of a semitone, while we predict musical shifts contained within one semitone.

\begin{figure*}[t]
\subfigure[]{\includegraphics[height=1.93in]{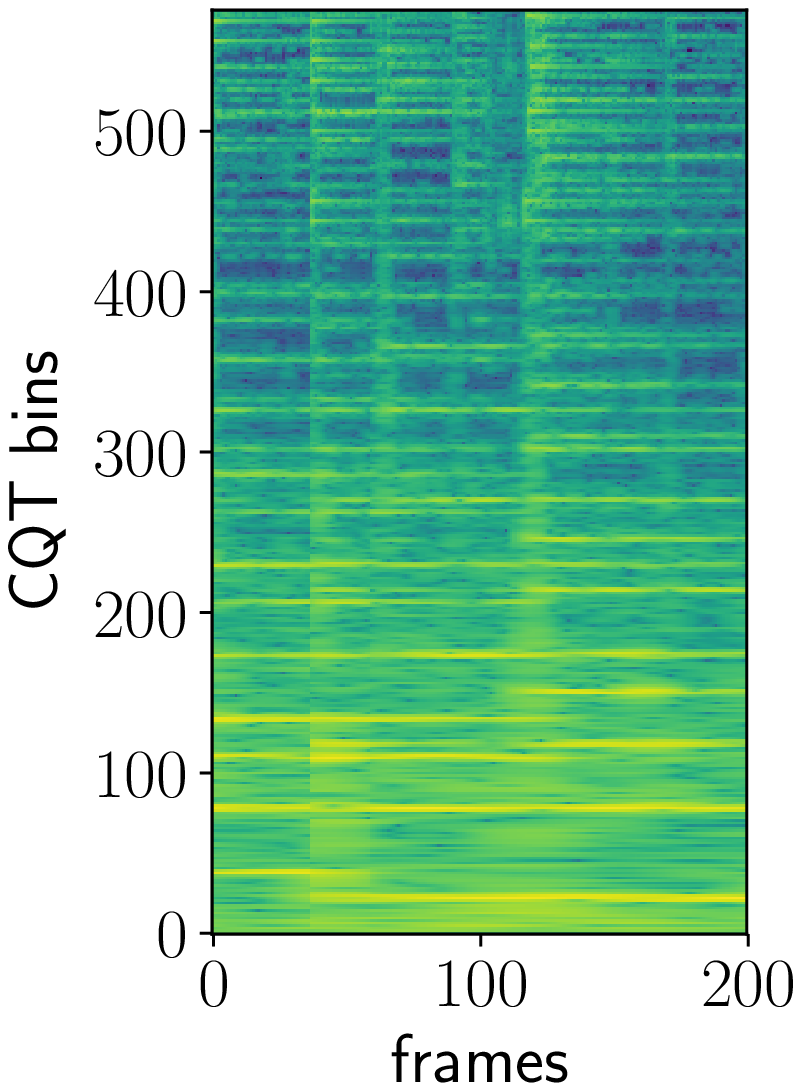}}
\subfigure[]{\includegraphics[height=1.93in]{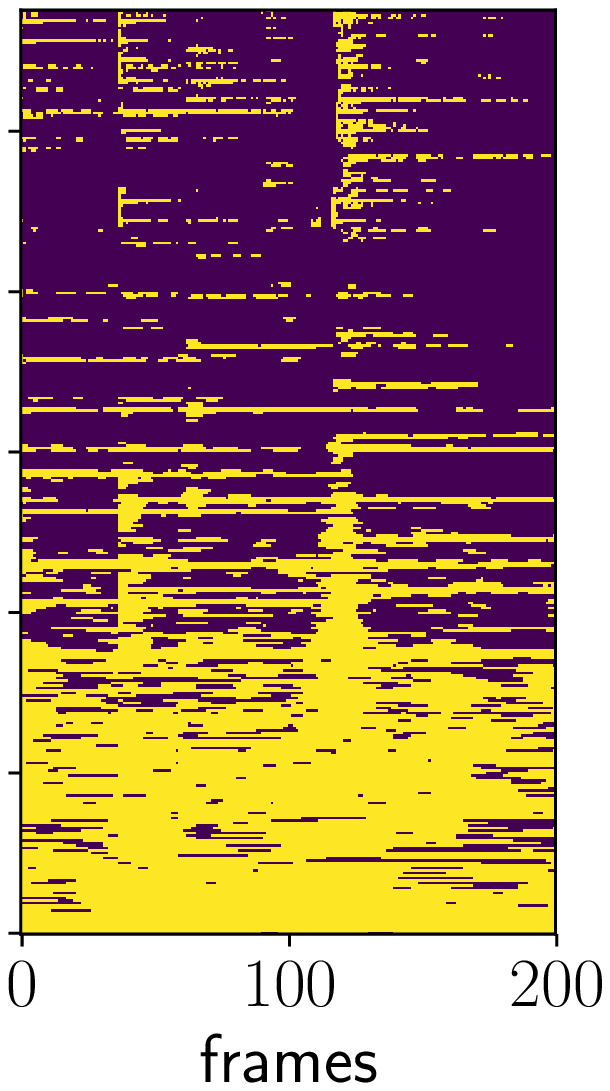}}
\subfigure[]{\includegraphics[height=1.93in]{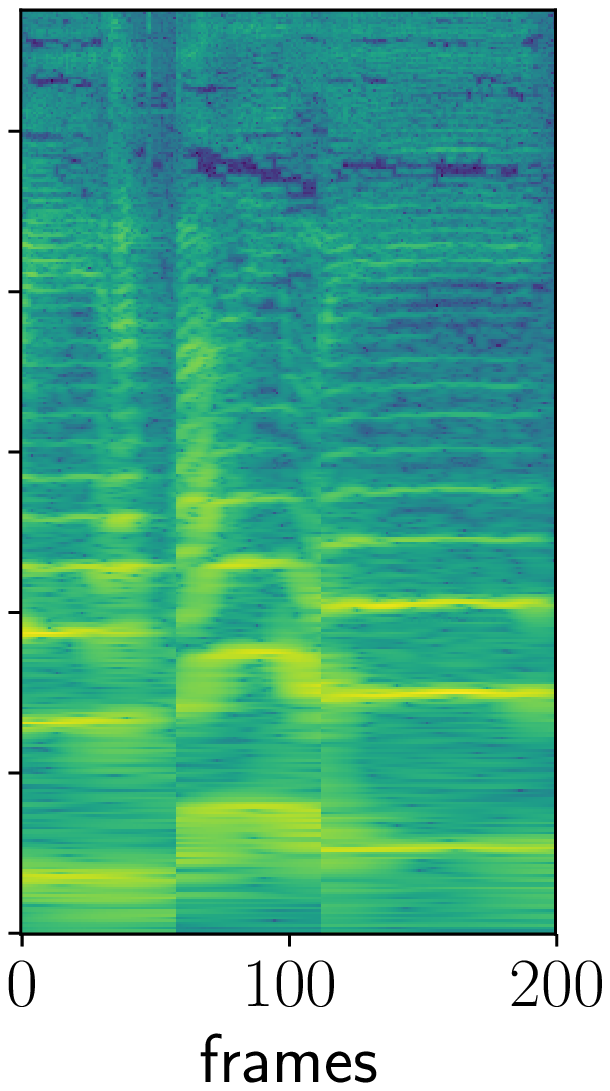}}
\subfigure[]{\includegraphics[height=1.93in]{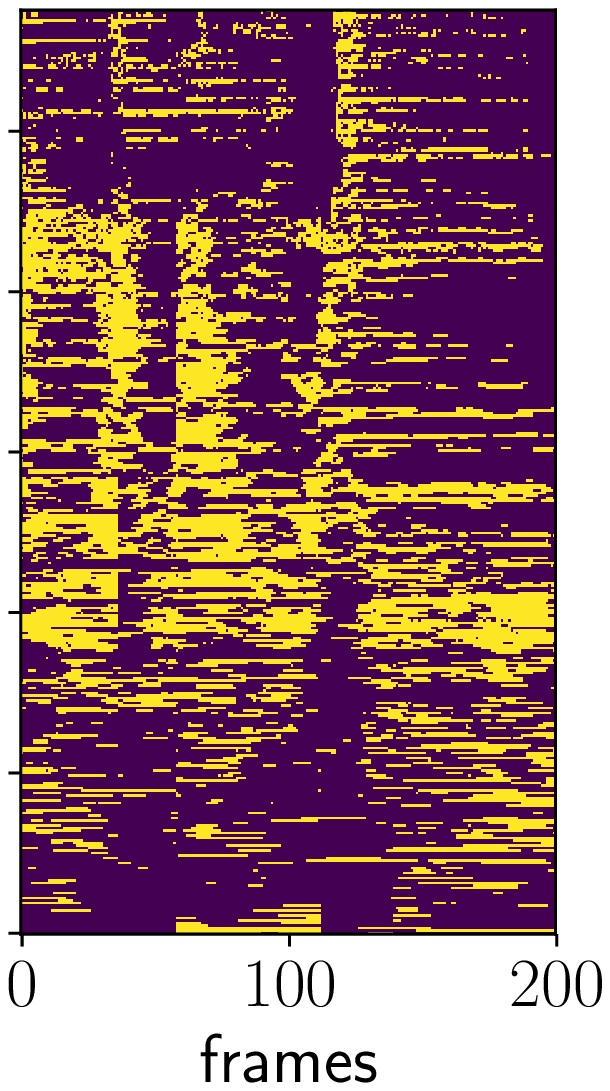}}
\subfigure[]{\includegraphics[height=1.93in]{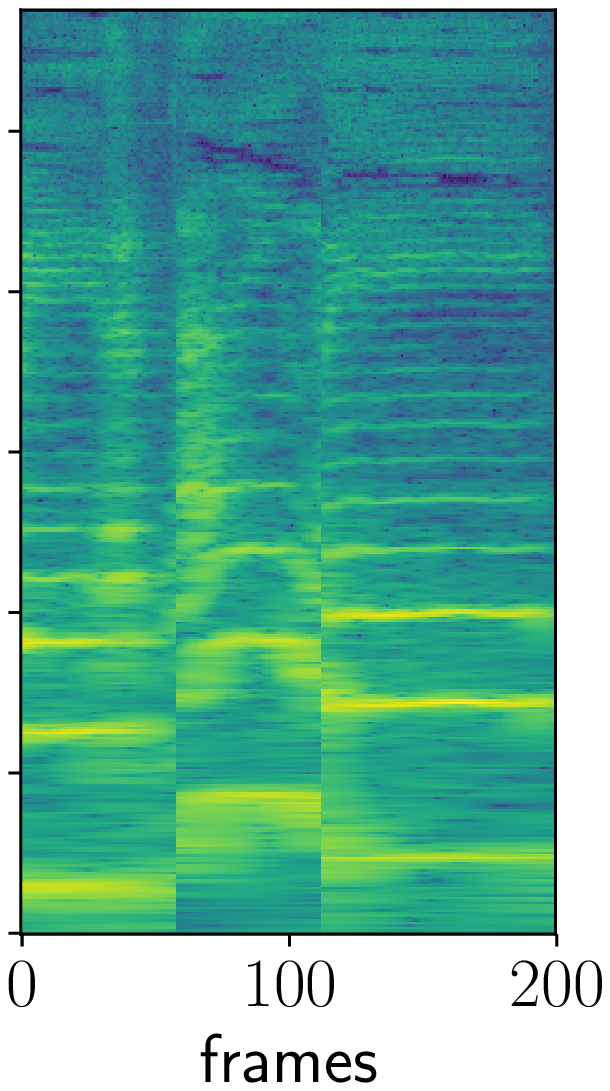}}
\subfigure[]{\includegraphics[height=1.93in]{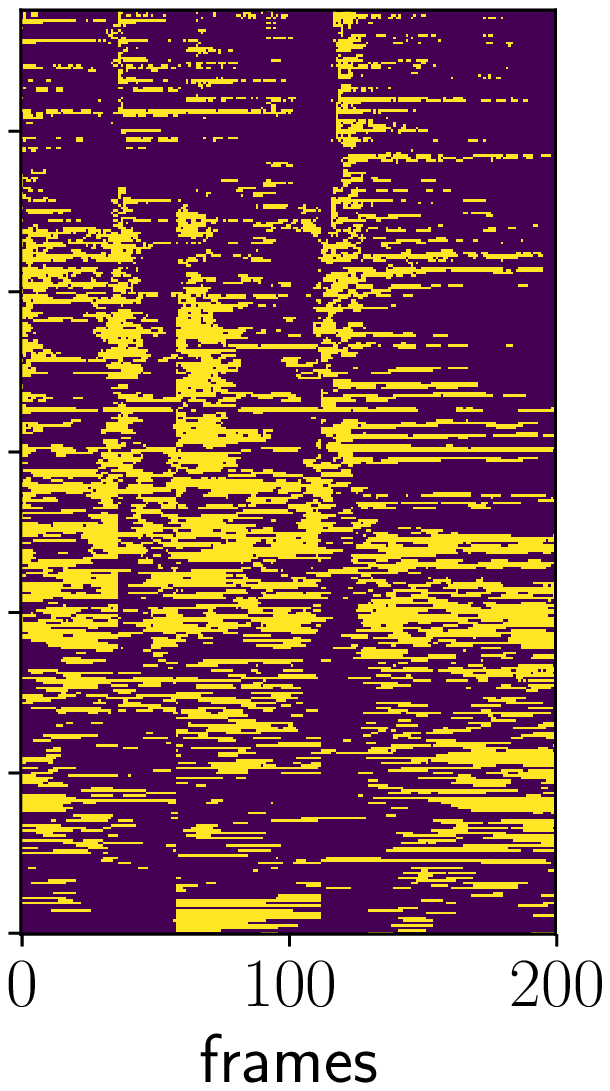}}
    \caption{Input data generation. The two first channels consist of the backing track and de-tuned vocals CQTs, shown in for three notes in (a) and (e). (c) has the original, in-tune vocals. The de-tuned vocals are shifted, by note, -80, 95, and -75 cents. A shift of less than a semitone is subtle but visible in the CQT. The third channel consists of the absolute value of the difference of the thresholded, binarized CQTs, shown in (f). (b) shows the binarized backing track: the same computation is applied to the vocal tracks. (d) shows the difference of binarized CQTs when the vocals were in tune. We expect that harmonics aligned with the backing track when audio is in tune will cancel out, where as misaligned harmonics will produce visible, straight lines. Some examples of this are found, for example, in the center right of (f) versus (d).}
    \label{fig:input}
\end{figure*}

\section{The Proposed Pitch Correction System}
\label{sec:model}
\subsection{Data}
We train the model with the ``Intonation" dataset \cite{wager2018intonation, damp2018}. The 4702 performances in the dataset were assembled from a larger database of Smule performances ranging from beginner to professional-level based on their tendency for good musical intonation, although they are not always perfectly in tune. It consists of 474 unique arrangements by 3556 singers. The performances are mostly of Western popular music. Given the karaoke nature of the performances, the monophonic vocal tracks are separate from the accompaniments. They were used in the ``raw" format, with no particular signal processing, e.g. denoising or filtering applied to them. We used one minute of audio from every performance for analysis, starting at 30 seconds into the recording. In order to generate training examples for the model, we use vocals that are de-tuned with various offsets with corresponding in-tune vocals. Timing and expressive gestures should be identical. We choose to generate out-of-tune singing by applying pitch shifts to in-tune singing while keeping the accompaniment fixed. After manually de-tuning the vocal tracks, we train the model to recover the de-tuning amounts. Smule artists' raw audio, including all pitch deviations, is treated as in-tune with intact expressive gestures. We use the simplifying assumption that a singer has one intended pitch for every note, and that the amount of de-tuning can only change at note boundaries, remaining otherwise constant. We deviate by only up to one semitone (100 cents) to avoid the ambiguity of choosing between semitones and larger intervals. The automatic Antares Auto-Tune has a similar scope as it centers the pitch around the nearest note. Our second simplifying assumption is that the amount of de-tuning between notes is independent.

For the proposed note-by-note data processing step, we find the note boundaries from the original singing performance, then apply the shift to the block of frames in the same note. The shifting is applied using Librosa's resampling and phase vocoder utilities \cite{mcfee2015librosa}. Every note in a melody is shifted by a constant value. We then generate 7 de-tuned versions of each song, where in every version, every note is shifted in either direction along the continuous logarithmic scale of cents by up to 100. The task of the network is to learn these shifts. In order to parse the melodies into notes for shifting, we analyzed the vocals pitch using the probabilistic Yin (pYIN) algorithm \cite{mauch2014pyin}. We could have treated frames where the pYIN analysis returned 0, meaning unvoiced or silent audio, to compute note boundaries. However, given that we had access to the musical scores, we aligned them to the pYIN analysis using Dynamic Time Warping (DTW) \cite{berndt1994using}. DTW computes a sequence of indices for both tracks that minimizes the total sum of distances between the two. We used the algorithm as described in \cite{muller2015fundamentals} and implemented in \cite{mcfee2015librosa}, setting the parameters to force most time distortions to be applied to the musical score, which consists of straight lines, instead of to the pYIN analysis. We then used the musical score to find the frame indices of note boundaries. Note that we used MIDI scores of the singing voice tracks only to detect the note boundaries, while the neural network does not take any score information as the input. We plan to eliminate this dependency on the note boundaries in future work by turning the algorithm into a frame-by-frame autotuner.

\subsection{Data format}
We convert the audio to the time-frequency domain, pre-processing the data to make its overtone frequencies evident. We compute the CQT of the vocals and accompaniment over 6 octaves with a resolution of 8 bins per semitone for a total dimension of 576. The lowest frequency is 100 Hertz. We use a frame size of 92 ms and a hop size of 11 ms. We normalize the audio by its standard deviation before computing the CQT. The vocals and accompaniment CQTs form two of the three input channels to the network. For the third channel, to contrast the difference between the first two channels, we binarize the two CQT spectrograms by setting each time-frequency bin to 0 if it has a value less than the global median of the song across all time and frequency, otherwise to 1. We then take the bitwise disagreement out of the two matrices based on the assumption that the in-tune singing voice will cancel out more harmonic peaks from the accompaniment track than the out-of-tune tracks. Figure \ref{fig:input} illustrates the data format and process of generating the bitwise disagreement.

\begin{table}[h!]
  \begin{center}
    \begin{tabular}{|c||c|c|c|c|}
    \hline
      & Conv1 & Conv2 & Conv3 & Conv4 \\
      \hline
      \#Filters/Units & 128 & 64 & 64 & 64 \\
      Filter size & (5, 5) & (5, 5) & (3, 3) & (3, 3) \\
      Stride & (1, 2) & (1, 2) & (2, 2) & (1, 1) \\
      Padding & (2, 2) & (2, 2) & (1, 1) & (1, 1) \\
      \hline
      & Conv5 & Conv6 & GRU & FC \\
      \hline
      \#Filters/Units & 8 & 1 & 64 & 1 \\
      Filter size & (48, 1) & (1, 1) & & \\
      Stride & (1, 1) & (1, 1) & & \\
      Padding & (24, 1) & (0, 0) & & \\
      \hline
    \end{tabular}
    \caption{The proposed network architecture}
    \label{tab:network}
  \end{center}
\end{table}

\subsection{Neural network structure}
A Convolutional Recurrent Neural Network (CRNN) allows the model to take context into account when predicting pitch correction, which is crucial for two reasons. First, the algorithm is expected to rely on aligning harmonics, which only occur in pitched sounds, but the vocal and backing tracks can both have unpitched or noisy sections. Second, a singer's note or melodic contour can last a second or multiple seconds and the choice of frequency at a given moment depends on the context. A recurrent structure in the model robustly keeps the memory of past activity and the pitched sounds of interest. The alignment would not always be obvious even for a musically proficient person if not for the long-term sequence analysis. We use convolutional filters to pre-process the three-dimensional data and reduce its dimensionality both in the time and frequency domains. The pre-processing makes it possible to have a single layer of GRU instead of a deeper one, which is computationally expensive and difficult to train. The GRU sequence length varies based on the number of frames in the note. The last output of the GRU is fed to a fully connected layer that predicts a single scalar output. We also keep track of the context by initializing the hidden state of every note with the previous note's final hidden state.

Table \ref{tab:network} displays the structure of the proposed network. Given that the input is a processed audio signal, its structure is different along the time and frequency axes, unlike images, which are translationally invariant. For this reason, we choose not to use max pooling, but use strides of two in the time axis in three convolutional layers. In the third layer, we also stride along the frequency axis, but perform this only once to compress without losing too much frequency information. The fifth convolutional layer has a filter of size 48 in the frequency domain, which maps to one octave in the CQT and captures frequency relationships in this larger space, as shown to be successful in \cite{bittner2017deep} and \cite{hsu2017learning}. The error function is the average Mean-Squared Error (MSE) between the pitch shift estimate and ground truth over the full sequence. We use the Adam optimizer \cite{kingma2014adam}, initialized with a learning rate of 0.00005. We do not use batch normalization because we only process one note at a time with seven differently shifted (but otherwise identical) versions. We apply gradient normalization \cite{pascanu2013difficulty} with a threshold of 100. The convolutional parameters were initialized using He \cite{he2015delving}, and the GRU hidden state of the first note of every song was initialized as a normal distribution with $\mu=0$ and $sd=0.0001$. The model was built in PyTorch \cite{paszke2017automatic}. To save processing time, after computing the random shifts and CQTs during the first epoch, we saved these to disk and shuffled the shifts at every note in following epochs.

\begin{figure}[t]
    \centering
    \includegraphics[width=0.5\textwidth]{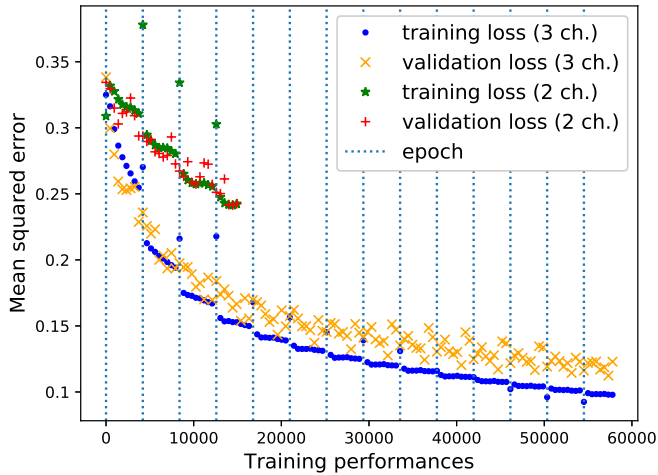}
    \caption{Training and validation losses when input data consists of either two or three channels. The loss was computed approximately nine times per epoch, every 500 performances, each of which consists of 50.4 notes (training samples) on average.}
    \label{fig:loss}
\end{figure}

\begin{figure}[t]
    \centering
    \includegraphics[width=0.5\textwidth]{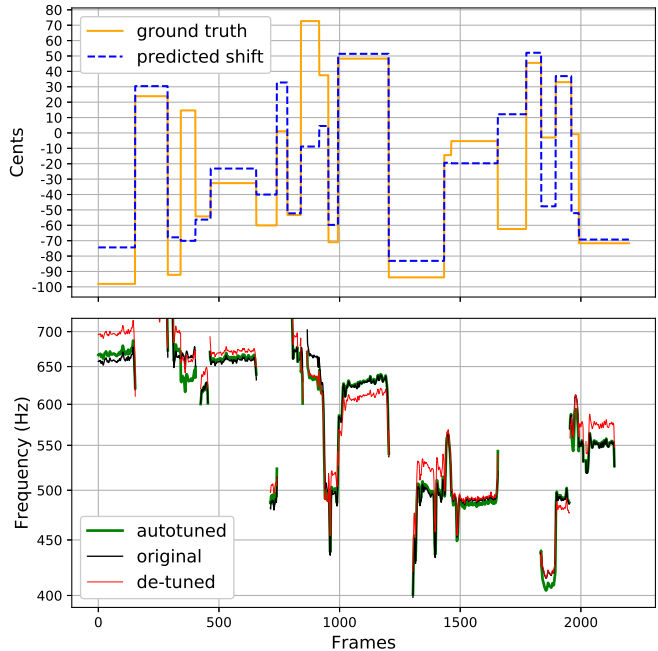}
    \caption{The upper plot shows predicted shifts and the ground truth for a validation performance. Every straight line segment shows the single scalar prediction for one note, stretched over the duration of frames. The bottom plot shows the original pitch track, which is considered in tune, the track after de-tuning, and the result of applying autotuning.}
    \label{fig:results}
\end{figure}

\section{Experimental Results and Discussion}
\label{sec:results}
We trained on the full dataset, excluding a validation list of 508 songs, selected not to share the same backing track as the training songs. Figure \ref{fig:loss} shows the training and validation loss, measured every 500 training songs, or 25200 notes on average. Within an epoch, training loss was computed as a running average over the samples visited up to that point within the epoch, while validation loss was computed every time over the full set. The average validation loss decreases to 0.123, corresponding to 35 cents. For comparison, we trained the model with two channels, omitting the thresholded difference channel. Figure \ref{fig:loss} shows that loss did not decrease as quickly. In part, this is because loss only decreased with a learning rate 10 times smaller. Figure \ref{fig:results} compares predicted shifts and the ground truth on a validation sample and shows the result of using these shifts to autotune the de-tuned pYIN pitch tracks. In many cases, the autotuned track is close to the original, but some notes are off by a large value. To autotune the singing voice, we input the predicted corrections to a phase vocoder or professional pitch shifting program.\footnote{Sample audio results are available at \url{http://homes.sice.indiana.edu/scwager/deepautotuner.html}} 

Although the proposed network learns the pitch shift of a given note of singing voice with a substantially low MSE, we have not tested it on real-world data where the out-of-tune singing voice might have different characteristics from the synthesized ones. To properly address this issue in the future, we plan to use a subjective test where users can listen to the outputs of this pitch correction algorithm and determine whether the singing sounds more in tune than before. When using an automatic approach without score, there will always be some errors where the correction is in the wrong direction. One way to address this is to combine the algorithm with user input or fall back to the Auto-Tune algorithm when necessary.

\section{Conclusion}
This experiment is the first iteration of a deep-learning model that estimates pitch correction for an out-of-tune monophonic input vocal track using the instrumental accompaniment track as reference. Our results on a CGRU indicate that spectral information the accompaniment and vocal tracks is useful for determining the amount of pitch correction required at a note level. This project is an initial prototype that we plan to develop into a model that predicts frame-by-frame corrections without relying on knowledge of note boundaries. The model can also be trained on different musical genres than Western music, especially genres with more fluidly varying pitch, when data is available. While the current model outputs the amount by which singing should be shifted in pitch, the model can easily be extended to perform autotuning, either by post-processing the voice recording or by developing the model to directly output the modified audio.


\vfill\pagebreak

% -------------------------------------------------------------------------
\bibliographystyle{IEEEbib}
\bibliography{strings.bib,refs}

\end{document}